\documentclass[journal]{IEEEtran}

\usepackage{cite}     
\usepackage{graphicx}  

\begin{document}

\title{Strong-field effects in the Rabi oscillations of the superconducting phase qubit}

\author{Frederick W. Strauch, S. K. Dutta, Hanhee Paik, T. A. Palomaki, K. Mitra, B. K. Cooper, R. M. Lewis, J. R. Anderson, A. J. Dragt, C. J. Lobb, and F. C. Wellstood
\thanks{Manuscript received August 30, 2006.
        This work was supported by the NRC, the NSF through the QuBIC Program, the NSA, and the Center for Superconductivity Research.}
\thanks{F. W. Strauch is with the National Institute of Standards and Technology,
Gaithersburg, Maryland 20899-8423 (email: frederick.strauch@nist.gov); S. K. Dutta, Hanhee Paik, T. A. Palomaki, K. Mitra, B. K. Cooper, R. M. Lewis, J. R. Anderson, A. J. Dragt, C. J. Lobb, and F. C. Wellstood are with the Center for Superconductivity Research and Joint Quantum Institute and the Department of Physics at the University of Maryland, College Park, Maryland 20742-4111 (email: sudeep@wam.umd.edu).}}


\maketitle

\begin{abstract}
Rabi oscillations have been observed in many superconducting devices, and represent prototypical logic operations for quantum bits (qubits) in a quantum computer. We use a three-level multiphoton analysis to understand the behavior of the superconducting phase qubit (current-biased Josephson junction) at high microwave drive power.  Analytical and numerical results for the ac Stark shift, single-photon Rabi frequency, and two-photon Rabi frequency are compared to measurements made on a dc SQUID phase qubit with Nb/AlO$_x$/Nb tunnel junctions.  Good agreement is found between theory and experiment.  
\end{abstract}

\begin{keywords}
Josephson junction, Rabi oscillation, multiphoton, qubit.
\end{keywords}


\section{Introduction}
\PARstart{S}{uperconducting} circuits utilizing Josephson junctions are a promising technology for quantum computation.  For the superconducting phase qubit, based on a single current- or flux-biased Josephson junction \cite{Ramos2001}, the spectroscopy of energy levels was first found some twenty years ago \cite{Martinis85}, while Rabi oscillations \cite{Martinis2002,Yu2002L}, multiphoton transitions \cite{Wallraff2003}, capacitive coupling \cite{Berkley2003}, and coupled oscillations \cite{Mcdermott2005} have only been demonstrated in the last five years.  While decoherence remains the dominant obstacle for these devices, a number of fundamental physical effects remain to be studied.

In this paper, we explore the physics of the phase qubit when irradiated by strong microwave fields.  A number of related experiments and calculations have appeared in the literature.  Multiphoton Rabi oscillations of a superconducting qubit were first performed in a charge qubit device \cite{Nakamura2001}, while the ac Stark shift was recently measured in a qubit-cavity system \cite{Schuster2005}.  High-power multiphoton spectroscopy of a three-junction flux qubit has been accomplished (with up to 20 photons) \cite{Oliver2005}.  The importance of diagonal terms in the microwave drive was theoretically explored for a SQUID qubit in \cite{AZafiris2005}, while many multiphoton and multilevel effects were recently studied with respect to EIT (electromagnetically-induced transparency) \cite{Dutton2006}. Three-level effects in the phase qubit have also been studied numerically \cite{Steffen2003}, perturbatively \cite{Meier2005}, and analytically \cite{Amin2006}.  

Here we compare analytical and numerical results from a three-level multiphoton analysis to Rabi oscillation experiments performed on a dc SQUID phase qubit.  We develop analytical results to describe the basic Rabi frequencies (the oscillation frequency between energy levels) and ac Stark shifts (the shift of the optimal resonance condition with applied power) of the device, which are in good agreement with experiment.  In addition to confirming the physics of the superconducting phase qubit, these results are important for situations where fast qubit operations are performed and for high-precision applications. 

\section{Model}
The Josephson phase qubit is simply a hysteretic current-biased Josephson junction that is well-isolated from dissipation in its leads \cite{Ramos2001}.  The Hamiltonian is
\begin{equation}
H = 4 E_c \hbar^{-2} p_{\gamma}^2 - E_J \left(\cos \gamma + (I/I_c) \gamma\right)
\end{equation}
where $\gamma$ is the gauge-invariant phase difference, $p_{\gamma}$ its conjugate momentum, $I_c$ the junction's critical current, $C$ its capacitance, $E_J = \hbar I_c / 2e$, and $E_c = e^2 / 2C$.  Here we consider a time-dependent bias current $I = I_{dc} - I_{ac} \cos \omega t$.  Key parameters for the system are the classical plasma frequency $\hbar \omega_0 = \sqrt{8 E_c E_J} \left(1 - (I_{dc}/I_c)^2 \right)^{1/4}$ and the barrier height (of the washboard potential) in units of $\hbar \omega_0$, denoted by $N_s$:
\begin{equation}
N_s = \frac{\Delta U}{\hbar \omega_0} \approx \frac{2^{3/4}}{3} \left(\frac{E_J}{E_c}\right)^{1/2} \left(1-\frac{I_{dc}}{I_c}\right)^{5/4},
\end{equation}
where the approximation is valid for $I_c-I_{dc} \ll I_c$.  In a three-level basis, the Hamiltonian has the matrix form
\begin{equation}
H = \left(\begin{array}{ccc} E_0 & 0 & 0 \\ 0 & E_1 & 0 \\ 0 & 0 & E_2 \end{array}\right) + A \cos \omega t \left(\begin{array}{ccc} x_{00} & x_{01} & x_{02} \\ x_{01} & x_{11} & x_{12} \\ x_{02} & x_{12} & x_{22} \end{array}\right),
\label{theory1}
\end{equation}
where we have ignored any overall energy shifts.  Using perturbation theory \cite{FWSThesis}, the energies are approximately
\begin{equation}
E_0 = \hbar \omega_0 \left(\frac{1}{2} - \frac{11}{8} \lambda^2 - \frac{465}{32} \lambda^4 \right),
\label{pt1}
\end{equation}
\begin{equation}
E_1 = \hbar \omega_0 \left(\frac{3}{2} - \frac{71}{8} \lambda^2 - \frac{5625}{32} \lambda^4 \right),
\label{pt2}
\end{equation}
\begin{equation}
E_2 = \hbar \omega_0 \left(\frac{5}{2} - \frac{191}{8} \lambda^2 - \frac{23475}{32} \lambda^4 \right),
\label{pt3}
\end{equation}
with $\lambda = (54 N_s)^{-1/2}$.  The (dimensionless) position matrix elements are, to order $\lambda^3$:
\begin{equation}
\begin{array}{llll}
x_{00} = & 3 \lambda/2, & x_{01} = & 2^{-1/2}(1 + 11 \lambda^2/4), \\
x_{11} = & 9 \lambda/2, & x_{12} = & 1 + 11 \lambda^2/2, \\
x_{02} = & -2^{-1/2}\lambda, & x_{22} = & 15 \lambda/2.  \\
\end{array}
\end{equation}
Finally, the drive amplitude has the explicit form
\begin{equation}
A = \frac{I_{ac}}{I_c}  E_J \left(\frac{8 E_c}{\hbar \omega_0}\right)^{1/2} \approx - \hbar \frac{d \omega_0}{d I} \frac{I_{ac}} {3 \lambda},
\end{equation}
and is linear in the microwave current $I_{ac}$.  

\section{Multiphoton Rotating Wave Approximation}
To describe multiphoton processes (see, {\it{e.g.}} \cite{Oliver2005,AZafiris2005}) we transform the Schr{{\"o}}dinger equation to an interaction picture using the time-dependent diagonal operator
\begin{equation}
H_0(t) = \left(\begin{array}{ccc} h_0 & 0 & 0 \\ 0 & h_1  & 0 \\ 0 & 0 & h_2\end{array}\right) + A \cos \omega t \left(\begin{array}{ccc} x_{00} & 0 & 0 \\ 0 & x_{11} & 0 \\ 0 & 0 & x_{22} \end{array}\right),
\end{equation}
where $h_0 = E_0$, $h_1 = E_0 + n \hbar \omega$, $h_2 = E_0 + 2 n \hbar \omega$, and $n$ is the number of photons involved in the $0 \to 1$ transition.  In the interaction-picture the new Hamiltonian is
\begin{equation}
\bar{H}(t) = e^{i \int_{0}^t H_0(s) ds /\hbar} H(t) e^{-i \int_{0}^t H_0(s) ds /\hbar} - H_0(t).
\end{equation}
By using the Bessel-function identity $e^{i z \sin \theta} = \sum_{m=-\infty}^{\infty} e^{i m \theta} J_{m}(z)$ and averaging $\bar{H}(t)$ over one period $T = 2\pi/\omega$ of the drive one obtains the multiphoton rotating wave Hamiltonian $H^{(n)}$:  
\begin{equation}
H^{(n)} = \hbar \left(\begin{array}{ccc} 0 & \Omega_{01}^{(n)}/2 & \Omega_{02}^{(n)}/2 \\ \Omega_{01}^{(n)}/2 & \epsilon_1^{(n)}  & \Omega_{12}^{(n)}/2 \\ \Omega_{02}^{(n)}/2 & \Omega_{12}^{(n)}/2 & \epsilon_2^{(n)} \end{array}\right)
\end{equation}
with $\epsilon_1^{(n)} = \omega_{01} - n \omega$, $\epsilon_2^{(n)} = \omega_{02} - 2 n \omega$, $\hbar \omega_{01} = E_1 - E_0$, $\hbar \omega_{02} = E_2 - E_0$, and
\begin{equation}
\Omega_{01}^{(n)} = \hbar^{-1} A x_{01} \sum_{s=0}^1 J_{n-1+2s} \left(\frac{A(x_{00}-x_{11})}{\hbar \omega}\right),
\end{equation}
\begin{equation}
\Omega_{02}^{(n)} = \hbar^{-1} A x_{02} \sum_{s=0}^1 J_{2n-1+2s} \left(\frac{A(x_{00}-x_{22})}{\hbar \omega}\right),
\end{equation}
\begin{equation}
\Omega_{12}^{(n)} = \hbar^{-1} A x_{12} \sum_{s=0}^1 J_{n-1+2s} \left(\frac{A(x_{11}-x_{22})}{\hbar \omega}\right).
\end{equation}

\section{Exact three-level analysis}
From the effective three-level Hamiltonian we can find exact solutions to the eigenvalue problem $H^{(n)} \psi = E \psi$.  Note that in this section we drop the $(n)$ dependence on the matrix elements.  Letting $E = \hbar x$, the exact eigevalues of $H^{(n)}$ are given by the roots of the cubic equation:
\begin{equation}
\begin{array}{l}
x^3 - (\epsilon_1 + \epsilon_2) x^2 + \left(\epsilon_1 \epsilon_2 - \frac{1}{4}(\Omega_{01}^2 + \Omega_{12}^2 + \Omega_{02}^2) \right) x \\
+ \frac{1}{4} (\epsilon_1 \Omega_{02}^2 + \epsilon_2 \Omega_{01}^2 - \Omega_{01} \Omega_{12} \Omega_{02}) = 0.
\end{array}
\label{hcubic}
\end{equation}
This has the exact solution (see also \cite{Amin2006}):
\begin{equation}
x_n = \frac{1}{3}(\epsilon_1 + \epsilon_2) + 2 p^{1/2} \cos(\theta + 2 n \pi/3)
\label{hcubic2}
\end{equation}
($n = 0, 1, 2$) with
\begin{equation}
\theta = \frac{1}{3} \arccos\left( - \frac{q}{2 p^{3/2}} \right),
\end{equation}
\begin{equation}
p =  \frac{1}{9}(\epsilon_1 + \epsilon_2)^2 - \frac{1}{3} \epsilon_1 \epsilon_2 + \frac{1}{12} (\Omega_{01}^2 + \Omega_{12}^2 + \Omega_{02}^2),
\end{equation}
and
\begin{equation}
\begin{array}{ll}
q =& -\frac{2}{27} (\epsilon_1 + \epsilon_2)^3 + \frac{1}{4} \left(\epsilon_1 \Omega_{02}^2 + \epsilon_2 \Omega_{01}^2 - \Omega_{01} \Omega_{02} \Omega_{12} \right) \\
& + \frac{1}{3} (\epsilon_1 +\epsilon_2) \left[ \epsilon_1 \epsilon_2 - \frac{1}{4} (\Omega_{01}^2 + \Omega_{12}^2 + \Omega_{02}^2) \right].
\end{array}
\end{equation}

The eigenvectors are
\begin{equation}
|v_n\rangle = N \left(\begin{array}{c} 2 (\epsilon_1 - x_n) \Omega_{02} - \Omega_{01} \Omega_{12} \\
-2 x_n \Omega_{12} - \Omega_{01} \Omega_{02} \\
\Omega_{01}^2 + 4 (\epsilon_1 - x_n) x_n \end{array}\right)
\end{equation}
with normalization
\begin{equation}
\begin{array}{lcl}
N &=& \left[ (2 (\epsilon_1 - x_n) \Omega_{02} - \Omega_{01} \Omega_{12})^2 \right. \\
&& + (-2 x_n \Omega_{12} - \Omega_{01} \Omega_{02})^2  \\
&& \left. + (\Omega_{01}^2 + 4 (\epsilon_1 -x_n)x_n)^2 \right]^{-1/2}.
\end{array}
\end{equation}
Using these eigenvectors one can calculate the time-dependent amplitudes for a given initial state.  That is, if 
\begin{equation}
|\psi(t)\rangle = a_0(t) |0\rangle + a_1(t) |1\rangle + a_2(t)|2\rangle
\end{equation}
then, given $a_n(0)$, the final amplitudes are
\begin{equation}
a_n(t) = \sum_{\ell, m=0}^2 e^{-i x_{\ell} t} \langle n |v_{\ell}\rangle \langle v_{\ell} | m\rangle a_m(0).
\end{equation}  

The eigenvalues $x_0$, $x_1$, and $x_2$ can be used to identify the various transitions, and the effective Rabi frequencies are given by the differences, {\it{e.g.}} $\bar{\Omega}_{R,01} = x_{0}-x_{2}$, $\bar{\Omega}_{R,02} = x_{2}-x_{1}$.

\section{Phase Qubit Approximations}
Here we describe approximations to the exact eigenvalues for the phase qubit, when $n=1$.  First, for typical drive amplitudes the Bessel functions can be approximated by $J_n(z) \approx (z/2)^n/n!$.  Then, using the position matrix elements of the cubic one finds
\begin{equation}
\Omega_{01}^{(1)} \approx \hbar^{-1} A x_{01}, \qquad \Omega_{12}^{(1)} \approx \sqrt{2} \Omega_{01}^{(1)},
\label{rabi1}
\end{equation}
and
\begin{equation}
\Omega_{02}^{(1)} \approx 3 \sqrt{2} \lambda^2 \frac{[\Omega_{01}^{(1)}]^2}{\omega}.
\end{equation}
We see that $\Omega_{02}^{(1)}$ is much smaller than $\Omega_{01}^{(1)}$ by two small factors: $\lambda^2$ and $\Omega_{01}^{(1)}/\omega$.  Thus, this matrix element can typically be neglected.

In this three-level approach, about each transition there will be an avoided crossing of two of the eigenvalues.  These avoided crossings indicate that oscillations between the original energy levels will occur, and with a frequency given by the eigenvalue difference.  This frequency can be approximated by using the characteristic Rabi formula such as 
\begin{equation}
\bar{\Omega}_{R,01} = \left[\Omega_{R,01}^2 + (\omega_{01} + \Delta \omega_{01} - \omega)^2\right]^{1/2}, 
\label{rabif}
\end{equation}
where $\Delta \omega_{01}$ is a possible shift of the resonance position due to the third level.  

The effect of this non-resonant level can be found perturbatively by either adiabatic elimination \cite{Dutton2006} or by a systematic expansion of the exact eigenvalues \cite{Amin2006}.  Here we simply state the final results.  For the $0 \to 1$ transition the minimum Rabi frequency occurs for
\begin{equation}
\omega_{01} = \omega - \Delta \omega_{01} \approx \omega - \frac{[\Omega_{01}^{(1)}]^2}{2 (\omega_{01}-\omega_{12})},
\label{stark1}
\end{equation}
thus identifying $\Delta \omega_{01} = \omega - \omega_{01}$.  This shift of the optimal transition point is known as the ac Stark shift in atomic physics, and has the characteristic dependence $\Omega^2/(\Delta \omega)$, where $\Delta \omega$ is the frequency difference of the off-resonant transition (here $\omega_{12}$) and the drive (here $\approx \omega_{01}$) and $\Omega$ the characteristic matrix element.  In addition, the minimal Rabi frequency is actually decreased from its nominal value of $\Omega_{01}^{(1)}$:
\begin{equation}
\Omega_{R,01}^{(1)} \approx \Omega_{01}^{(1)} \left(1 - \frac{[\Omega_{01}^{(1)}]^2}{4 (\omega_{01}-\omega_{12})^2} \right).
\label{rabi2}
\end{equation}

The two-photon $0\to2$ transition, when $\Omega_{02}^{(1)}$ can be neglected, is produced by two off-resonant single-photon transitions $0\to1\to2$, with a Rabi frequency of 
\begin{equation}
\Omega_{R,02}^{(1)} \approx \frac{\sqrt{2} [\Omega_{01}^{(1)}]^2}{(\omega_{01} - \omega_{12})}.
\label{rabi3}
\end{equation} 
In addition, a similar Stark shift is found in the $0 \to 2$ transition, with the resonance condition:
\begin{equation}
\omega_{02}/2 \approx \omega + \frac{[\Omega_{01}^{(1)}]^2}{4(\omega_{01}-\omega_{12})}.
\label{stark2}
\end{equation}  
Note that in (\ref{stark1}) and (\ref{stark2}) the transition frequencies appear on both sides of the equation, and thus should be determined self-consistently; this is typically a small effect.  

For these expressions to be valid, the condition $\Omega_{01}^{(1)} \ll \omega_{01} - \omega_{12}$ must be met.  While this is not the case for the highest power results of the following section, these perturbative results do indicate the main physics.  In the following, we will compare both these and the full three-level multiphoton effects predicted by (\ref{hcubic2}) with experiment.

\begin{figure}
\centering
\includegraphics[width=3in]{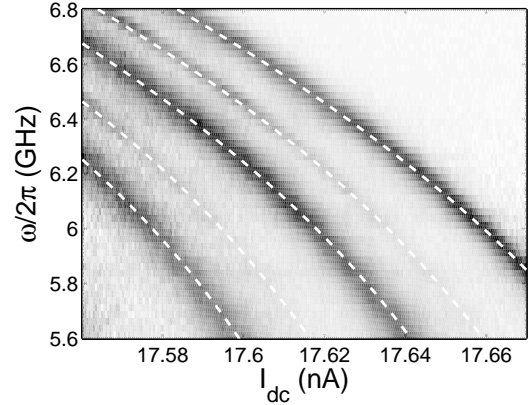}
\caption{Experimental microwave spectroscopy of a Josephson phase qubit, scanned in frequency (vertical) and bias current (horizontal).  Dark points indicate experimental microwave enhancement of the tunneling escape rate, while white dashed lines are quantum mechanical calculations of (from right to left) $\omega_{01}$, $\omega_{02}/2$, $\omega_{12}$, $\omega_{13}/2$, and $\omega_{23}$.}
\label{fig1}
\end{figure}

\begin{figure}
\centering
\includegraphics[width=3in]{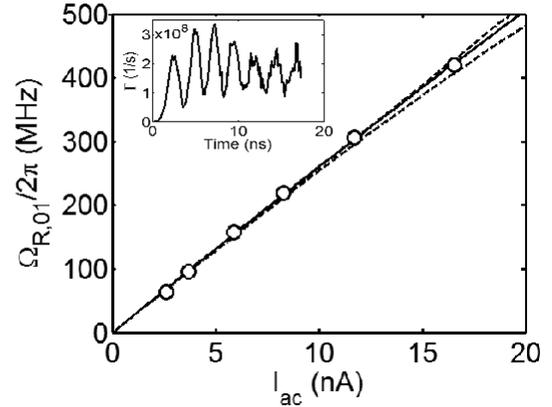}
\caption{Rabi frequency $\Omega_{R,01}$ of the one-photon $0\to1$ transition as function of microwave current $I_{ac}$.  The dots are experimental data, the solid line predictions from the three-level model, and the dashed lines are the lowest-order results (\ref{rabi1}) (top) and second-order (\ref{rabi2}) (bottom) perturbative results.  The inset shows Rabi oscillations of the escape rate for $I_{ac} = 16.5$ nA.}
\label{fig2}
\end{figure}

\begin{figure}
\centering
\includegraphics[width=3in]{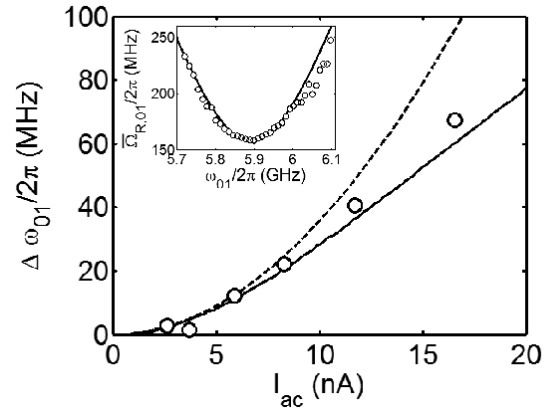}
\caption{The ac Stark shift $\Delta \omega_{01}$ of the one-photon $0\to1$ transition as function of microwave current $I_{ac}$. The dots are experimental data, the solid line predictions from the three-level model, and the dashed line perturbative results.  The inset shows the oscillation frequency $\bar{\Omega}_{R,01}$ as a function of the level spacing $\omega_{01}$ for $I_{ac} = 5.87$ nA and the fit using (\ref{rabif}) to obtain $\Omega_{R,01}$ and $\Delta \omega_{01}$.}
\label{fig3}
\end{figure}

\begin{figure}
\centering
\includegraphics[width=3in]{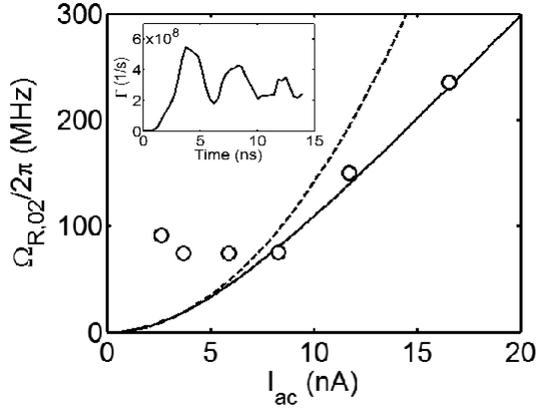}
\caption{Rabi frequency $\Omega_{R,02}$ of the two-photon $0\to2$ transition as function of microwave current $I_{ac}$.  The dots are experimental data, the solid line predictions from the three-level model, and the dashed line perturbative results.  Inset shows Rabi oscillations of the escape rate for $I_{ac} = 16.5$ nA.}
\label{fig4}
\end{figure}

\section{Experiment}

Measurements were made on a dc SQUID phase qubit with two Nb/AlO$_x$/Nb Josephson junctions in a 6-turn 6 nH loop.  The devices were fabricated at Hypres, Inc., using a trilayer process \footnote{Mention of commercial products or services in this paper does not imply approval or endorsement by NIST, nor does it imply that such products or services are necessarily the best available for the purpose.} and the qubit junction had an area of $(10\mu\textrm{m})^2$. The SQUID was dynamically initialized to one of its metastable wells and simultaneously current- and flux-biased in such a way to study the properties of one of its junctions; further experimental details can be found in \cite{Palomaki2006}.  Microwave spectroscopy \cite{Martinis85} performed on the device yielded single and two-photon transitions in quantitative agreement with quantum mechanical spectra of a single junction with parameters $C = 4.46$ pF and $I_c = 17.821$ $\mu$A; a typical scan in frequency and bias current is shown in Fig.~1.  Applying microwaves of constant frequency $\omega/2\pi = 5.9$ GHz, resonances were seen for bias currents matching a single or two-photon transition.  Near $I_{dc} = 17.660$ $\mu$A the single-photon $0\to1$ transition was observed, corresponding to energy level spacings of $\omega_{01}/2\pi = 5.9$ GHz and $\omega_{12}/2\pi = 4.99$ GHz.  For the same microwave frequency but at a smaller bias $I_{dc} = 17.636$ $\mu$A the two-photon $0\to2$ transition was observed, with levels spacings of $\omega_{01}/2\pi =6.23$ GHz and $\omega_{12}/2\pi = 5.57$ GHz.  

About each transition, and for several microwave powers, Rabi oscillations were observed in the tunneling rate of the junction from the zero to finite voltage state \cite{Yu2002L} (see inset to Fig.~2).  The Rabi oscillation frequency, shown in Fig.~2, was found by varying the bias current and noting the level spacing at which maximal oscillations of the escape rate and minimal Rabi frequency was located (see inset to Fig.~3).  Also shown are the exact three-level results and approximate results of the previous section.  To produce these curves a single fitting parameter was used to calibrate the applied microwave current $I_{ac}$ as a function of the applied microwave power $P$.  This calibration (specifically, $I_{ac} =  58.7 \, {\textrm{nA}} (P/{\textrm{mW}})^{1/2}$) was then used for the horizontal axis.  Note also that the numerical values for $\omega_{01}$ and $\omega_{12}$ were used in the theory, as the perturbative results (\ref{pt1})-(\ref{pt3}) are accurate only for $N_s > 3$ \cite{FWSThesis} while here $N_s \approx 2.21$.  

Using this calibration, the ac Stark shift $\Delta \omega_{01} = \omega - \omega_{01}$ (derived from the bias current) is plotted in Fig.~3.  The solid curve is the three-level prediction, with no fitting parameters, while the dashed curve is the perturbative result (\ref{stark1}).  We see good agreement with the three-level results, even for powers where the perturbative result fails. 

Finally, the Rabi frequency of the two-photon $0\to2$ transition is shown in Fig.~4.  Here the solid curve is the three-level prediction while the dashed curve is the approximation (\ref{rabi3}).  For small powers, the experimental results are likely dominated by systematic errors in the tunneling rate measurements.  For large powers we find that while the perturbative result fails, the full three-level prediction is in excellent quantitative agreement with the data.

\section{Conclusion}
In conclusion, we have examined the behavior of the superconducting phase qubit when irradiated with microwaves at high power.  We have identified two effects that are new to this qubit: multiphoton Rabi oscillations and the ac Stark shifts.  We have also shown that a simple three-level theory can explain, with minimal fitting parameters, both of these subtle experimental effects.  This is quite remarkable, as dissipation, decoherence, and tunneling are all signficant in the same parameter regime for these devices.  Finally, these results indicate that high-power Rabi oscillations can in principle be used to characterize all of the relevant matrix elements needed for high-fidelity microwave control of superconducting phase qubits.



\end{document}